\documentstyle[sprocl,epsf]{article}

\def\myitem#1 {\hangindent\parindent\indent\llap{#1)\enspace}\ignorespaces}

\def\E{\,\rlap/\!E_T}  

\newcommand{\lsim}{\mathrel{\raisebox{-.6ex}{$\stackrel{\textstyle<}{\sim}$}}}
\newcommand{\gsim}{\mathrel{\raisebox{-.6ex}{$\stackrel{\textstyle>}{\sim}$}}}

\catcode`@=11
\def\@cite#1#2{\unskip\nobreak\relax
    \def\@tempa{$\m@th^{\hbox{\the\scriptfont0 #1}}$}%
    \futurelet\@tempc\@citexx}
\def\@citexx{\ifx.\@tempc\let\@tempd=\@citepunct\else
    \ifx,\@tempc\let\@tempd=\@citepunct\else
    \let\@tempd=\@tempa\fi\fi\@tempd}
\def\@citepunct{\@tempc\edef\@sf{\spacefactor=\the\spacefactor\relax}\@tempa
    \@sf\@gobble}
\catcode`@=12

\begin{document}

\font\fortssbx=cmssbx10 scaled \magstep1
\hbox to \hsize{
\hbox{\fortssbx University of Wisconsin - Madison}
\hfill$\vcenter{\normalsize\hbox{\bf MADPH-98-1034}
                \hbox{January 1998}
                \hbox{\hfil}}$ }

\bigskip

\title{Supersymmetry Phenomenology at Hadron Colliders\footnote{Talk presented at {\it COSMO 97: International Workshop on Particle Physics and the Early Universe}, Lancaster, UK, Sept.~1997.}}

\author{V. Barger}

\address{Physics Department, University of Wisconsin, Madison, Wisconsin 53706, USA}

\maketitle

\thispagestyle{empty}

\abstracts{The phenomenology of a low-energy supersymmetry at hadron colliders is discussed with consideration of the minimal supergravity model, with a large top quark Yukawa coupling at the grand unification mass scale, and gauge mediated symmetry breaking models. Possible supersymmetry interpretations of some unexplained events are mentioned.}

\section{Introduction}

Supersymmetry has many possible faces depending on the nature of the lightest supersymmetric particle (LSP), the relative sparticle masses, and whether $R$-parity is conserved. In the minimal supersymmetric model (MSSM) each standard model fermion (boson) has a boson (fermion) superpartner; see Table~1.
\begin{table}[h]
\caption{Particle content of the MSSM for one generation of fermions}
\medskip
\epsfxsize=2.9in
\centering\leavevmode
\epsffile{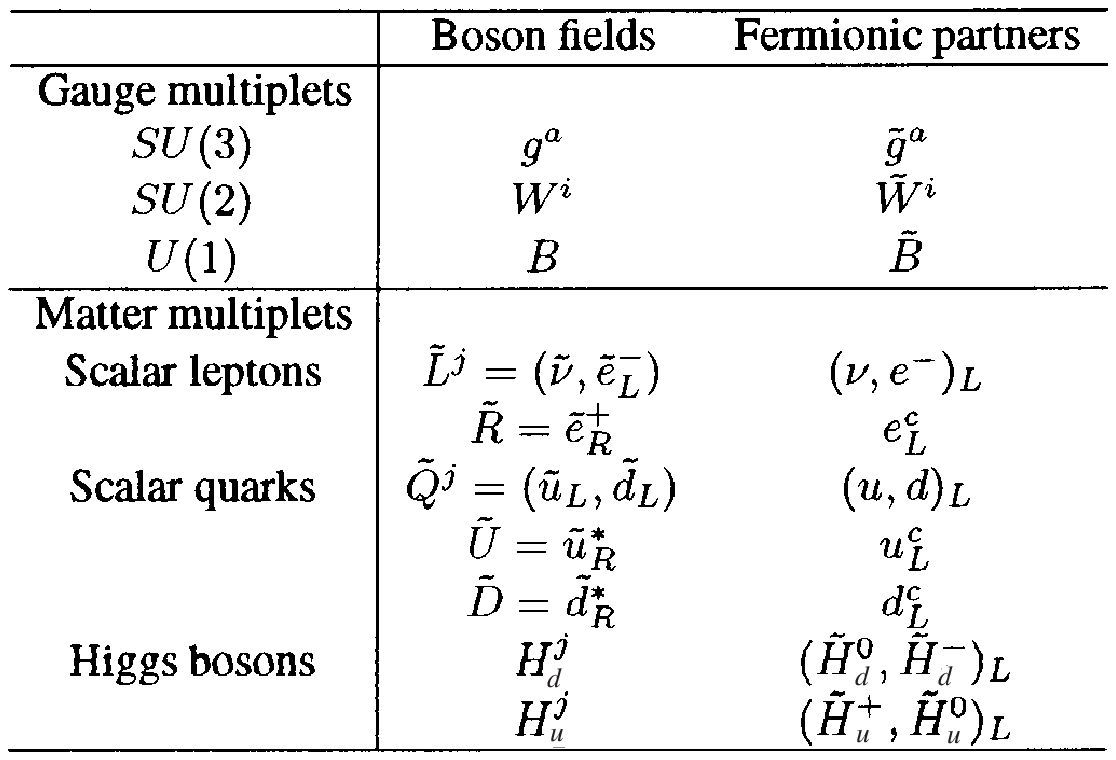}
\end{table}
 Supersymmetry breaking occurs via ``soft'' masses ($m_{1/2}$, $m_0$, $\mu$) and couplings ($A_{ijk}, B$); in the generic MSSM there are over 100 of these soft parameters. An $R$-parity is defined by the relation $R=(-)^{3(B-L)+2S}$, where $B$, $L$, and $S$ denote baryon number, lepton number, and spin; $R=+1$ for Standard Model (SM) particles and $R=-1$ for the supersymmetric (SUSY) partners. The nature of the phenomenology depends on whether $R$ is conserved or not --- baryon number violating or lepton number violating interactions are okay, but not both, because of proton stability.
With $R$ non-conservation, the LSP is unstable. With $R$ conservation, there are various stable LSP options: $\tilde B$ (gaugino), $\tilde H$ (higgsino), $\tilde G$ (gravitino), $\tilde g$ (gluino). The $\tilde B$  option is particularly interesting for the LSP as the cold dark matter candidate.

SUSY breaking models are based on mediation from a hidden sector to the visible sector. The models give relations among the soft parameters and thus have predictive power.  The two classes of models are minimal SuperGravity (mSUGRA) and gauge mediated symmetry breaking (GMSB). In mSUGRA, the breaking occurs at a high scale $\sim10^{10}$~TeV via gravitational interactions which give universal soft terms (with possible Planckian non-universal corrections to the scalar masses). In GMSB the breaking typically takes place at an effective scale $\Lambda \sim 10$--100~TeV via messenger fields. The dominant part of the sparticle masses in GMSB models are obtained via gauge interactions.

This review presents a general overview of supersymmetry phenomenology at present and future hadron colliders. Other more extensive recent reviews\cite{tata-susy,reviews,gmsb-physrep} can be consulted for references to the vast primary literature.

\section{mSUGRA Phenomenology}

The renormalization group equations relate masses and couplings at the scale of the Grand Unified Theory (GUT) to their electroweak scale values. The effective parameter set is the GUT scale masses $m_{1/2}$ and $m_0$, the GUT scale trilinear coupling $A_0$, sign($\mu$), and $\tan\beta$ where $\tan\beta = v_u/v_d$ is the ratio of the vacuum expectation values of the Higgs bosons that give masses to the up-type quarks and leptons and the down-type quarks and leptons. There are a number of generic predictions that are a consequence of a large top-quark Yukawa coupling at the GUT scale:

\renewcommand{\labelenumi}{\theenumi)}
\begin{enumerate}

\item  The Higgs miracle happens at $M_Z$; electroweak symmetry breaking (EWSB) occurs radiatively.

\item  $m_b/m_\tau$ is correctly predicted from $\lambda_b=\lambda_\tau$ unification at the GUT scale.

\item $\lambda_t$ has an infrared fixed point, predicting
{\arraycolsep=.1em
\begin{eqnarray}
\tan\beta &\simeq& 1.8 \ \ (m_t = 200{\rm~GeV} \, \sin\beta)\nonumber\\
&\mbox{or}&\nonumber\\
\tan\beta &\simeq& 56 \nonumber
\end{eqnarray}}%

\item Gaugino masses at scale $M_Z$ are given by
\begin{eqnarray}
&& M_1/\alpha_1 = M_2/\alpha_2 = M_3/\alpha_3 \nonumber\\
&& M_1 = 0.44 m_{1/2},\ M_2=0.88 m_{1/2},\ M_3 = 3.2 m_{1/2} \nonumber
\end{eqnarray}

\item $|\mu|$ is large compared to $M_1, M_2$ at scale $M_Z$.

\item The chargino mass matrix has the approximate form
\[ {\cal M} \sim \left( \begin{array}{cc}
M_2 & 0 \\ 0 & -\mu
\end{array} \right) \mbox{ in the }
\left( \begin{array}{c} \tilde W^\pm\\ \tilde H^\pm \end{array} \right)\rm\ basis  \]
Thus $\tilde\chi_1^\pm\sim\tilde W^\pm$ and $\tilde\chi_2^\pm\sim\tilde H^\pm$. 

\item The neutralino mass matrix has the approximate form 
\[ {\cal M} = \left( \begin{array}{cccc}
M_1& 0 \\ 0& M_2 \\ & & 0& \mu\\ & & \mu& 0 \end{array} \right)
\mbox{ in the } \left( \begin{array}{c} 
\tilde B\\ \tilde W^3\\ \tilde H^0_d\\ \tilde H^0_u \end{array} \right) \rm\ basis \]
Thus $\tilde\chi_1^0 \sim \tilde B^0$ and $\tilde\chi_2^0\sim \tilde W^3$.

\item The sparticle mass ratios are approximately
\[ \tilde\chi_1^0 : \tilde\chi_2^0 : \tilde\chi_1^\pm : g = 1:2:2:7 \]
$\tilde\chi_1^0$ is the LSP.

\item The relic density of dark matter\cite{ellis} is explained by the LSP for parameters\cite{ourdm}
\[\begin{array}{lcl}
 m_0 \lsim 200{\rm~GeV},& 80\lsim m_{1/2} \lsim 450{\rm~GeV} & \rm for\ \tan\beta\sim 1.8\\
 m_0 \gsim 300{\rm~GeV},& 500\lsim m_{1/2} \lsim 800{\rm~GeV} & \rm for\ \tan\beta\sim50
\end{array}\]

\item There is an upper bound on the mass of the lightest MSSM Higgs boson of 
$m_{h^0} \lsim M_Z |\cos2\beta|$ at tree level and $m_{h^0} \lsim 130$~GeV including radiative corrections.

\item The other Higgs bosons have masses
\[ \begin{array}{lll}
m_A \approx m_H \gg M_Z& \rm if& \tan\beta\sim1.8\\
m_A \sim{\cal O}(M_Z)& \rm if& \tan\beta\gsim m_t/m_b
\mbox{ or }m_0\sim50\rm~GeV
\end{array} \]

\item The colored particles (squarks and the gluinos) are heavier; the scalar masses depend on $m_0$	
\[ \tilde\chi_1^0, \tilde\chi_2^0, \tilde\chi_1^\pm, \tilde\ell, h \mbox{ are ``light''} \]
\[ \tilde\chi_3^0, \tilde\chi_4^0, \tilde\chi_2^\pm, \tilde g, \tilde q \mbox{ are ``heavy''} \]

\item The stop mass-squared matrix has the form
\[ \tilde m^2 = \left( \begin{array}{cc}
L^2& am_t\\ am_t& R^2 \end{array}\right) \mbox{ in the }
\left( \begin{array}{c} \tilde t_L\\ \tilde t_R \end{array} \right)\rm\ basis\]
where $a=A_t + \mu\cot\beta$. The mixing may be large and consequently $\tilde t_1$ may be lighter than the other squarks. 
\end{enumerate}

\section{Classic mSUGRA Signatures}

The primary signatures for supersymmetry at hadron colliders are (i) missing transverse energy ($\E )+{}$jets, (ii) same-sign dileptons${}+\E+{}$jets, and (iii) trileptons${}+\E$. Gluinos and squarks are strongly produced in pairs ($\tilde g\tilde g,\ \tilde g\tilde q,\ \tilde q\tilde q$). These sparticles decay through multistep cascades. An example is gluino decay through the chain
\begin{equation}
\tilde g\to q\tilde q,\quad \tilde q\to q\tilde\chi_i,\quad \tilde\chi_i\to \tilde\chi_j W,\quad \tilde\chi_j\to f\bar f\tilde\chi_i^0 \,.
\end{equation}
The $\tilde\chi_1^0$ is a source of $\E$. Leptons result from chargino and heavier neutralino decays. Since the gluino is a Majorana particle, the decay rates for $\tilde g\to\tilde\chi^+$ and $\tilde g\to\tilde\chi^-$ are the same. Consequently, $\tilde g\tilde g$ and $\tilde g\tilde q$ production lead to same-sign dilepton events. The process $q\bar q\to\tilde\chi_1^+\tilde\chi_2^0$ with the decays $\tilde\chi_1^+\to\tilde\chi_1^0\ell^+\nu$ and $\tilde\chi_2^0\to\tilde\chi_1^0 \ell^+\ell^-$ gives trilepton events. 

Tevatron searches by the CDF and D\rlap/0 collaborations\cite{culbert} exclude low mass gluino and squarks, as follows, for the representative case of $m_{\tilde g} = m_{\tilde q}$ and typical choices of other SUSY parameters (e.g., $\tan\beta=2$, $A_0=0$):

\begin{tabular}{lccc}
&&& $m_{\tilde g} = m_{\tilde q}$\\
channel& expt.& luminosity& excluded\\
$\E +{}$jets& D\rlap/0& 80 pb$^{-1}$& $\lsim260$ GeV\\
$\E +{}$dielectrons& D\rlap/0& 93 pb$^{-1}$& $\lsim 267$ GeV\\
$\E +{}$same sign dileptons + 2 jets& CDF& 90 pb$^{-1}$& $\lsim 225$ GeV
\end{tabular}

\noindent
CDF and D\rlap/0 limits from the $\E +{}$trileptons channel on the lightest chargino mass are not as restrictive as those obtained\cite{kats} at LEP\,2. However, at the upgraded Tevatron, trilepton events will provide the highest SUSY mass reach, up to $m_{\chi_1^\pm}\sim230$~GeV with 10~fb$^{-1}$ luminosity.

Stop searches have been made in $t\bar t$ events, where one top decays via $t\to bW$ and the other decays via the SUSY modes $t\to \tilde t_1\tilde\chi_1^0$ with $\tilde t_1\to c\tilde\chi_1^0$ or $\tilde t_1\to b\tilde\chi_1^+\to bq\bar q'\tilde\chi_1^0$ or $b\ell^+\nu\tilde\chi_1^0$, using the fact that the jets in SUSY modes are softer than the jets in SM events. The region $m_{\tilde t_1}\lsim95$~GeV is excluded by the combination of CDF\cite{culbert} data and the LEP\,2 excluded region $m_{\tilde\chi_1^+}\lsim90$~GeV\cite{kats}. 

Three ``odd'' $t\bar t$ events have been found by CDF and D\rlap/0 in the channel $\ell^+\ell^- jj\E$, which have characteristics that are unlike or improbable of those expected from $t\bar t$. A proposed explanation is cascade decays of squarks\cite{barnett}
\begin{equation}
\tilde q\to\tilde\chi_1,\quad \tilde\chi\to\nu\tilde\ell,\quad \tilde\ell\to\ell \tilde\chi_1^0 \,.
\end{equation}
The production rate and kinematics for these events suggest masses 
\begin{eqnarray*}
&& m_{\tilde q_L}\sim310{\rm\ GeV},
\quad m_{\tilde\chi^+} = m_{\tilde\chi_2^0}\sim 260{\rm\ GeV},\\
&& m_{\tilde\ell} \sim210\mbox{--}230{\rm\ GeV},\quad m_{\tilde\chi_1^0} \sim 50\rm\ GeV \,.
\end{eqnarray*}
The inferred $m_{\tilde\chi_2^0}/m_{\tilde\chi_1^0}$ ratio is not consistent with mSUGRA. 

The discovery reach for SUSY particles at future hadron colliders is summarized\cite{tata-susy} in Table~2. Note that the LHC reach for sleptons is only $\sim300$~GeV due to the small Drell-Yan signal at high mass and to large SM backgrounds.

\begin{table}[h]
\caption[]{Discovery reach of future colliders.\cite{tata-susy}}
\epsfxsize=4.3in
\centering\leavevmode
\epsffile{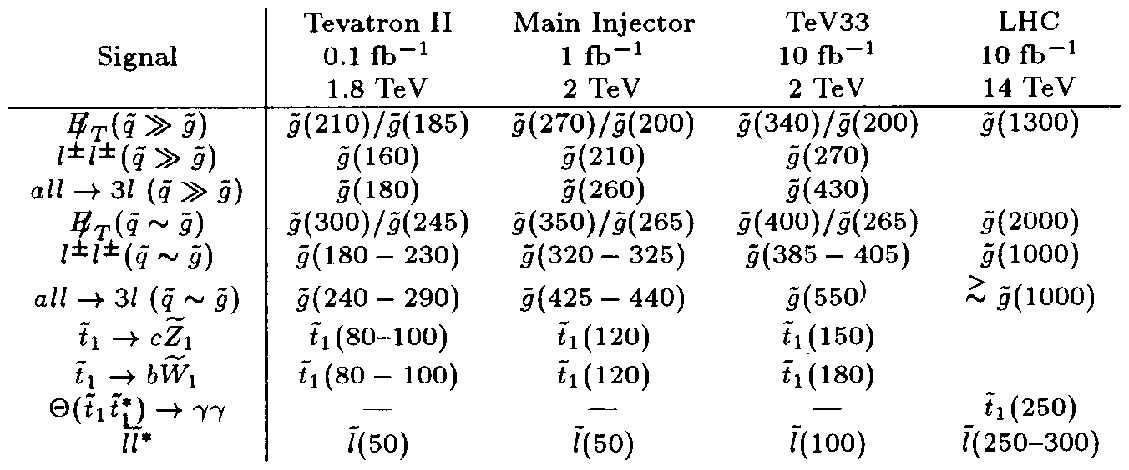}
\end{table}

At the LHC a first estimate of the SUSY mass scale is the peak of the distribution in the quantity\cite{hinch}
\begin{equation}
M_{\rm eff} = \sum_{i=1}^4 \left| p_{T\,{\rm jet}, i} \right| + \E \,.
\end{equation}
This peak measures
\begin{equation}
M_{\rm SUSY} \equiv \min (m_{\tilde g}, m_{\tilde q})
\end{equation}
to an accuracy of $\pm10\%$. The complexity of gluino and squark cascade decays makes general simulations of sparticle mass reconstruction difficult at the LHC, but specific channels have been analyzed with considerable success. The effective strategy is to identify characteristic signatures of particles near the end of the cascade decays as a starting point and then work back up the decay chain. A sample demonstration is $\tilde g\to b\tilde b$, $\tilde b\to b\tilde\chi_2^0$, $\tilde\chi_2^0\to\ell^+\ell^-\tilde\chi_1^0$. The signal for $\tilde g\tilde g$ events is then $\geq 2$ $b$ jets and $\geq1$ $\ell^+\ell^-$ pair. The endpoint of the $\ell^+\ell^-$ invariant mass distribution measures $M_{\tilde\chi_2^0}-M_{\tilde\chi_1^0}$ to $\pm50$~MeV. From this, $m_{1/2}$ is inferred and then $M_{\tilde\chi_1^0}$ and $M_{\tilde g}$ can be predicted by mSUGRA relations. Using this $M_{\chi_1^0}$ the events can be reconstructed and checks can be made for consistency. From the reconstructed $m_{\tilde b}$ the value of $m_0$ can be deduced. In this particular case the overall fit to the simulation compared to the input was\cite{hinch}
\begin{eqnarray*}
m_0 &=& 200^{+13}_{-\phantom08}\rm\ GeV\ \ (200\ GeV)\\
m_{1/2} &=& 99.9 \pm 0.7\rm\ GeV\ \ (100\ GeV)\\
\tan\beta &=& 1.99\pm0.05\ \ (2)
\end{eqnarray*}
and the sign of $\mu$ was correctly deduced. 

\section{GMSB Phenomenology}

In gauge mediated SUSY breaking models the gravitino is the LSP, with mass $m_{\tilde G} \sim {\cal O}$(keV). The soft masses are determined from one input scale $\Lambda$ (other parameters are $\tan\beta$, a vacuum expectation value $F$ and the messenger mass scale $M_{\rm messenger}$). The gaugino mass spectrum is the same as mSUGRA, with
\begin{equation}
M_i(\mu) \sim {\alpha_i(\mu)\over 4\pi} \Lambda
\end{equation}
at one loop, but the scalar mass spectrum is different:
\begin{equation}
m^2(\Lambda) \sim \Lambda^2 \sum c_i \alpha_i^2(\Lambda) + D\mbox{-}terms \,.
\end{equation}

The candidates for the next-to-lightest supersymmetric particle (NLSP) are $\tilde\chi_1^0$, $\tilde\ell_R$, and $\tilde \tau_1$, with decays
\begin{equation}
\tilde\chi_1^0\to \gamma \tilde G,\ \tilde\ell_R \to \ell \tilde G(\ell=e,\mu),\ \tilde\tau_1\to \tilde G\tau\,.
\end{equation}
These decays may occur within the detector or outside, depending on the mass and energy of the NLSP and the value of $F$. Decays of more than one NLSP may contribute (co-NLSP).
The SUSY particles are produced in pairs and then undergo chain decays to the NLSPs wich decay to $\tilde G$. The possible signals are summarized\cite{ambros,kolda} in Table~3. 

\begin{table}[t]
\caption[]{Signals of GMSB models.\cite{kolda}}
\bigskip
\epsfxsize=2.5in
\centering\leavevmode
\epsffile{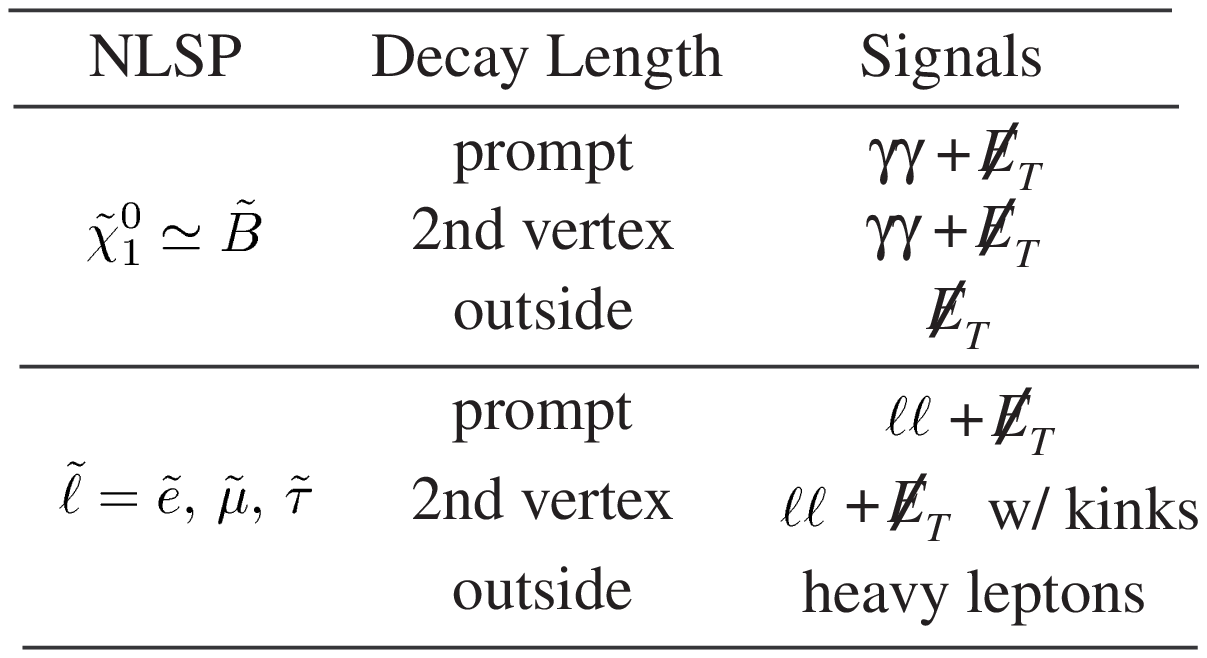}
\end{table}

The CDF $2e+2\gamma+\E$ event with hard gammas and electrons has two possible GMSB interpretations:\cite{ambros,kolda,dimop}
\begin{enumerate}
\item $p\bar p\to\tilde e\tilde e\to e\tilde\chi_1^0\to ee\gamma\gamma\tilde G\tilde G$,
\item $p\bar p\to\tilde\chi_1^+\tilde\chi_1^-\to e\nu\tilde\chi_1^0\to ee\gamma\gamma\nu\nu \tilde G\tilde G$.
\end{enumerate}
However, these interpretations are disfavored by the prediction\cite{dimop,dimop2} of numerous related events from
\begin{equation}
p\bar p\to \tilde\chi_1^+\tilde\chi_1^-, \tilde\chi_1^+\tilde\chi_2^0, \tilde e\tilde e, \tilde\nu\tilde\nu
\end{equation}
that are not observed. Thus the lone CDF $ee\gamma\gamma \E$ event remains an enigma. Niether D\rlap/0 nor CDF find any other diphoton events with large $\E$ and leptons. A search for diphotons\cite{culbert} with $\E > 25$~GeV by D\rlap/0\cite{culbert} excludes chargino masses $m_{\chi_1^\pm}<156$~GeV in gauge mediated models.

Some GMSB models have been proposed that have unusual features. In multi-scale models\cite{ambros2} the first two generations of squarks and sleptons may have masses up to 40~TeV; because they are weakly coupled to Higgs bosons the $<1$~TeV naturalness requirement may be evaded. In a model constructed by Raby\cite{raby} the gluino is light ($\lsim100$~GeV) and stable, but the rest of the sparticle spectrum resembles mSUGRA; in this model there are no $\E$ signals from gluinos. In these model variants the signatures of SUSY could differ significantly from 
more standard scenarios. However, in all models $\tilde\chi_1^0$, $\tilde\chi_1^\pm$, and $\tilde\chi_2^0$ are expected to be light, so gauginos are secure targets for experimental searches.

\section{SUSY Higgs Searches at Hadron Colliders}

In the decoupling limit, which applies in mSUGRA for low $\tan\beta$, the lightest SUSY Higgs boson ($h_{\rm MSSM}$) has coupling strengths that are similar to the SM Higgs boson ($h_{\rm SM}$). For decoupling, the prospects for $h_{\rm MSSM}$ discovery are essentially the same as for $h_{\rm SM}$. With the main injector the Tevatron should have the capability to detect $h_{\rm SM}$ up to $\sim120$~GeV via $Wh_{\rm SM}$ production with $h_{\rm SM}\to b\bar b$ decay\cite{kim}. This mass reach almost covers the interesting range for $h_{\rm MSSM}$. The LHC can cover the full range for $h_{\rm MSSM}$ and likely also find\cite{bk2} the other SUSY Higgs bosons $H^0$, $A^0$ and $H^\pm$.

\section{HERA Anomaly and SUSY $R$-parity Violation}

The H1 and ZEUS collaborations initially observed an excess over SM expectations of neutral current events in $e^+p$ deep inelastic scattering\cite{adloff}. However, after further running, the effect has become less compelling. Table~4 summarizes the present situation in both neutral current (NC) and charged current (CC) channels\cite{altar}. The H1 NC data show an enhancement in the $e^++{}$jet invariant mass bin $M=200\pm12.5$~GeV; see Table~5. This could be an indication of an $s$-channel resonance in the NC channel or it could just be a statistical fluctuation.

\begin{table}[h]
\epsfxsize=2.5in
\caption[]{HERA high-$Q^2$ events ($Q^2\gsim1.5\times10^4\rm~GeV^2$) and SM expectations.\cite{altar}}
\bigskip
\centering\leavevmode
\epsffile{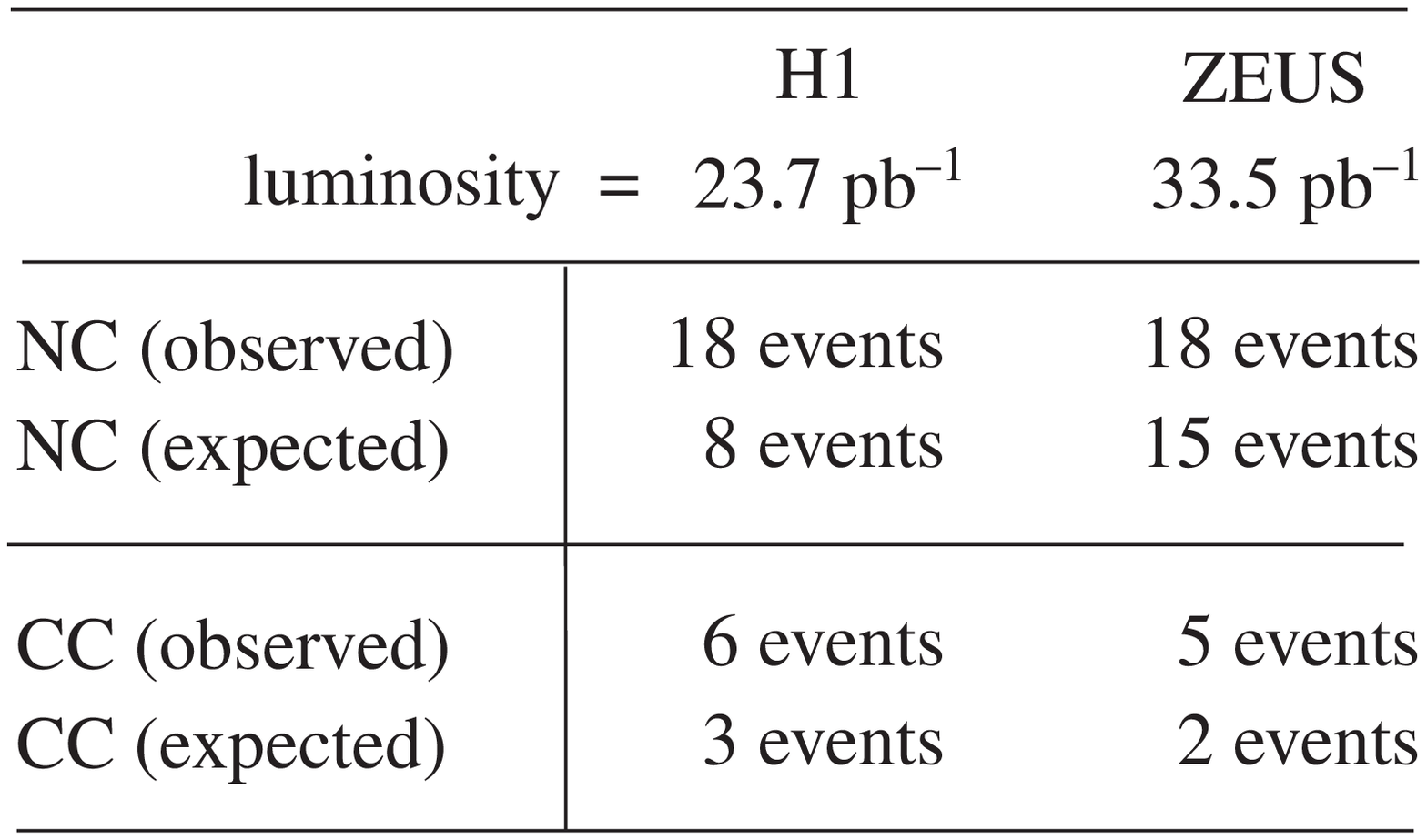}
\end{table}

\begin{table}[t]
\epsfxsize=2.5in
\caption[]{HERA high-$Q^2$ events  in the $e^+p$ invariant mass bin $M=200\pm12.5$~GeV.}
\bigskip
\centering\leavevmode
\epsffile{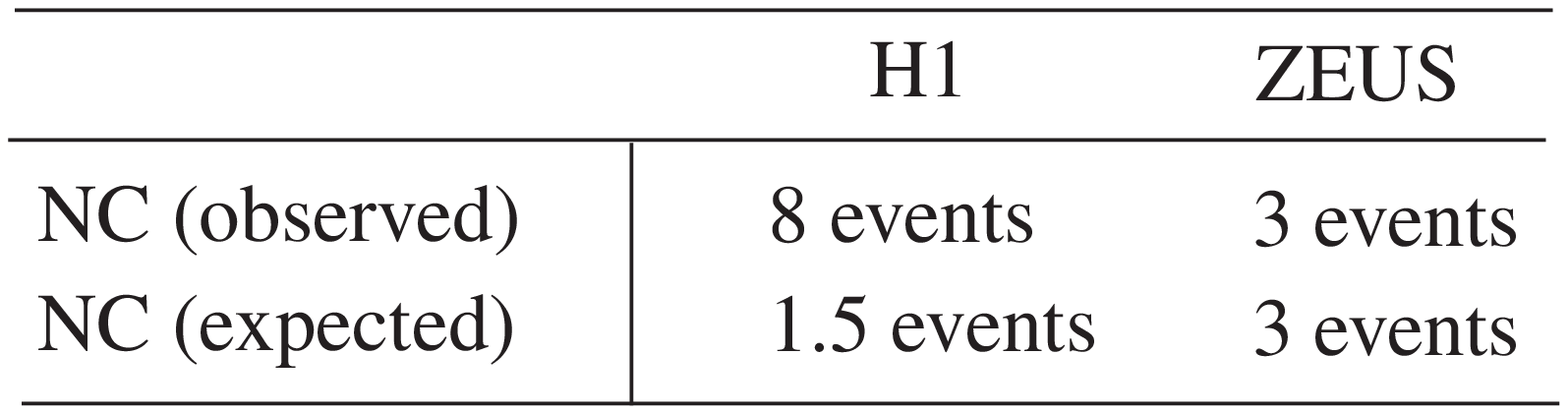}
\end{table}

With SUSY $R$-parity violating (RPV) couplings, the stop could contribute as a resonance in the $e^+p$ $s$-channel. For the $e^+_R s_R\to \tilde t_L\to e^+s$ process the anomalous NC event rate observed by H1 could be explained with $\lambda' \sqrt B \sim 0.15$ to 0.25, where $\lambda'$ is the $es\tilde t$ coupling and $B$ is the $\tilde t\to e^+s$ branching fraction; for $e^+d\to \tilde t_L\to e^+d$, $\lambda'\sqrt B\sim 0.025$ to 0.033 would be needed\cite{altar}. Tevatron searches\cite{schell} for $e^+$ jet events place the 95\% CL limits $m_{\tilde t_1}>225$~GeV (D\rlap/0) and $m_{\tilde t_1}>240$~GeV (CDF) assuming a branching fraction $B=1$. A branching fraction $B(\tilde t_1\to e^+s)\lsim 0.5$--0.7 is thus needed for compatiblity of the Tevatron limits and the H1 NC event excess. The competing $\tilde t_1\to b\tilde\chi^+$ $R$-conserving decay could give the required branching fraction. Carena et al.\cite{carena} suggested that the decay chain $\tilde t_1\to b\tilde\chi^+,\ \tilde\chi^+\to c\bar{\tilde b},\ \bar{\tilde b} \to \nu \bar d$ could explain the excess CC events (with the same $\lambda'$ coupling as for the NC anomaly) if the $\tilde t_1$, $\tilde\chi^+$ and $\tilde b$ masses were closesly spaced so that the $b$ and $c$ are not observed as jets. Thus, an $s$-channel stop resonance with RPV couplings could give high $Q^2$ NC and CC events {\em but} the case for new physics at HERA is not yet compelling.
The possibility of explaining HERA high-$Q^2$ events with contact interactions is limited\cite{last-bchz} because of constraints from other data, especially Drell-Yan lepton-pair production at the Tevatron. 

\section{Summary}

The Tevatron and LHC colliders can do ``the job'' --- find the particles of low energy supersymmetry, measure their masses (with the exception of the LSP mass), determine their properites {\em or} reject the hypothesis. The Tevatron is now placing interesting mass limits on SUSY particles. In the mSUGRA model the present Tevatron bounds are $m_{\tilde g} = m_{\tilde q} \gsim 267$~GeV and $m_{\tilde t_1} \gsim 95$~GeV. In GMSB models with $\tilde\chi^0\to\gamma\tilde G$ the chargino mass bound is $m_{\tilde\chi^+}\gsim150$~GeV. Some unexplained events could be of SUSY origin --- the CDF ``odd" $tt$ events, the CDF $\gamma\gamma ee\E$ event --- but there is not a convincing smoking gun yet.

With the advent of GMSB models many SUSY options abound and the search for supersymmetry has become correspondingly more complex. The mSUGRA mdoel has a fully satisfactory candidate for Cold Dark Matter --- the lightest neutralino. In GMSB models a sneutrino messenger of mass $\cal O$(TeV) is a possible CDM particle\cite{hanhempf}.

\section*{Acknowledgments}

I am grateful to Chung Kao for helpful advice in the preparation of this report. This research was supported in part by the U.S.~Department of Energy under Grant No.~DE-FG02-95ER40896 and in part by the University of Wisconsin Research Committee with funds granted by the Wisconsin Alumni Research Foundation.

\end{document}